# MANIFESTATION OF THE STRONG QUADRUPOLE LIGHT-MOLECULE INTERACTION IN THE SEHR SPECTRA OF PHENAZINE AND PYRAZINE

**A.M. Polubotko**

A.F. Ioffe Physico-Technical Institute Russian Academy of Sciences, Politechnicheskaya 26, 194021 Saint Petersburg, Russia. Fax: 297-10-17, E-mail: alex.marina@mail.ioffe.ru

The paper demonstrates possibility of giant enhancement of Surface Enhanced Hyper Raman Scattering on the base of qualitative consideration of electromagnetic field near some models of rough metal surfaces and of some features of the dipole and quadrupole light-molecule interaction, such as it was made in the dipole-quadrupole SERS theory. Consideration of symmetrical molecules allows to obtain selection rules for their SEHR spectra and establish such regularity as appearance of the lines, caused by totally symmetric vibrations, transforming after the unit irreducible representation in molecules with the groups, where the elements of symmetry change $z$ on $-z$. These lines are forbidden in usual HRS spectra. Analysis of literature data on phenazine and pyrazine molecules demonstrates that their SEHR spectra can be explained by the SEHRS dipole-quadrupole theory. Analysis of the SEHR spectra of these molecules reveals appearance of strong forbidden bands, caused by vibrations transforming after the unit irreducible representation that strongly confirms this theory which allows to interpret the whole SEHR spectra in detail. The results corroborate this common mechanism of Surface Enhanced optical processes on molecules adsorbed on rough metal surfaces.

# 1. Introduction

Understanding of mechanism of surface enhancement of optical processes by molecules adsorbed on rough metal surfaces is very important. However in spite of there are some reliable conceptions, which strongly corroborate our point of view, the mechanism is still a matter of debates. There are several theories and approaches, which try to explain these phenomena. The most widespread one is the approach, which is based on the conception of the enhancement by surface plasmons, which may be excited on a metal surface. One should note that plasmons can not be excited on flat surfaces at the applied incident frequencies. Therefore one usually connects their appearance with existence of surface roughness. The conception of the surface plasmons is well defined on flat metal surfaces [1]. However it becomes indefinite on a random rough surface, which exists in real experiments. In addition (in case we suppose existence of such excitations) the whole field, which affects the molecule, is determined as a sum of all surface and other modes, existing near rough surface. Therefore we have a great doubt in the fact that some separate plasmon modes determine the whole giant enhancement.

The second conception is the charge transfer theory, or the charge transfer enhancement mechanism [2] which explains these processes by resonance Raman scattering arising between metal electron states and some unoccupied molecular states, which are named as affinity levels. The main objection against this mechanism is that it occurs only in presence of surface roughness. It is quite ununderstandable, why this mechanism is not observed on single metal surfaces that was strongly confirmed in numerous experiments of Allan Campion (see [3-6] for example). The second objection is that this mechanism must be observed on some definite frequencies, corresponding to the electron transfer from the electronic metal states to the unoccupied states of the molecule. Then it must occur on the frequencies, which depend on the type of the molecule, while the frequency dependence of the enhancement in SERS is not resonant and SERS is observed practically for all molecules in a very large frequency range on arbitrary rough surfaces [7]. One should note, that the charge transfer enhancement mechanism arose from the experimental fact, that the enhancement in the first layer of adsorbed molecules, which have direct contact with the surface is significantly stronger than the enhancement in the second and upper layers. One should note, that this fact was explained in [8-10]. It was demonstrated, that the first layer effect is caused by a very large difference in the values of the electric field and their derivatives in the vicinity of prominent places with a large curvature just in the first and the second layers of adsorbed molecules and has a pure electrodynamical but not the chemical nature [8-10]. The third and the most important opinion on the nature of the enhancement is that the enhancement is caused by so-called rod effect, or by increase of the electric field near the roughness of the wedge, or tip or cone, or spike type [11-17]. The enhancement at the tops of these ideal features is infinity. Therefore this fact guarantees the strong enhancement near the tops of the roughness of the wedge or rod, or spike like forms. This fact well corroborates the main reason of SERS and other surface enhanced processes-the strong surface roughness. In addition this mechanism was well corroborated by experiments of Emory and Nie [18, 19]. However it should be noted, that the pure rod effect is not able to explain such features of the SERS, SEHRS [7, 20-22] and probably SERRS spectra, as appearance of the strong forbidden bands in molecules with the groups, where the elements of symmetry change $z$ on $-z$. Sometimes this fact is explained by distortion of molecules due to adsorption, however we consider that it is caused by existence of so called strong quadrupole light-molecule interaction and the actual reason of the surface enhanced processes are the dipole and especially quadrupole light-molecule interactions arising in surface fields strongly varying in space in the vicinity of the sharp surface roughness [10, 17]. Moreover, since we usually deal with random surfaces and random surface electromagnetic fields, the values of amplitudes of separate bands, which can change in a broad range, can not be well defined characteristics, which can well establish the dipole-quadrupole scattering mechanism. The most reliable informative characteristics, which allow to make a unequivocal conclusion about the enhancement mechanism, are the irreducible representations of the vibrations, which cause forbidden and allowed lines and other features in the enhanced spectra of symmetrical molecules. Analysis of the SERS spectra of symmetrical molecules [10, 15, 17, 23-26] strongly confirms the dipole-quadrupole mechanism of SERS. Here we shall demonstrate, that the main available regularities of the SEHR spectra of phenazine and pyrazine, which refer to symmetrical molecules with $D_{2h}$ symmetry group strongly confirm our point of view and the validity of the dipole-quadrupole enhancement mechanism, arising due to the strong rod effect. One should note that our approach, based on such models of the surface roughness as wedge, tip, cone or spike [10-13,15-17] is used in a new branch of enhanced spectroscopy- Tip Enhanced Raman spectroscopy (TERS) [27,28]. The singular behavior near the tops of these models explains existence of very strong local optical fields associated with the features of surfaces on the tops of these models, which are responsible for the giant enhancement. One should note, that there can be some another configurations and models of roughness, such as two close parallel round nanowires [29] where the strong enhancement occurs in the gap between them. However in any case the enhancement in our and the other models occurs in some local areas and leads to the enhancement of the electric fields and its derivatives and hence to increase of the dipole and quadrupole light-molecule interactions in these regions.

## 2. Electromagnetic field near rough metal surface

The main property of electromagnetic field near rough metal surface is its strong spatial heterogeneity. As an example we can consider a model of the rough surface - a strongly jagged metal lattice with a regular triangular profile (Figure 1). The electromagnetic field above this lattice can be represented in the form

$$\overline{E} = \overline{E}_{inc} + \overline{E}_{surf..sc} \qquad (1)$$

where,

$$\overline{E}_{inc} = \overline{E}_{0,inc} e^{-ik_0 \cos\theta_0 z + ik_0 \sin\theta_0 y}$$
$$\left|\overline{E}_{0,inc}\right| = 1 \qquad (2)$$

$\theta_0$ - is the angle of incidence,

$\overline{k}_0$ is the wave vector of the incident field in a free space,

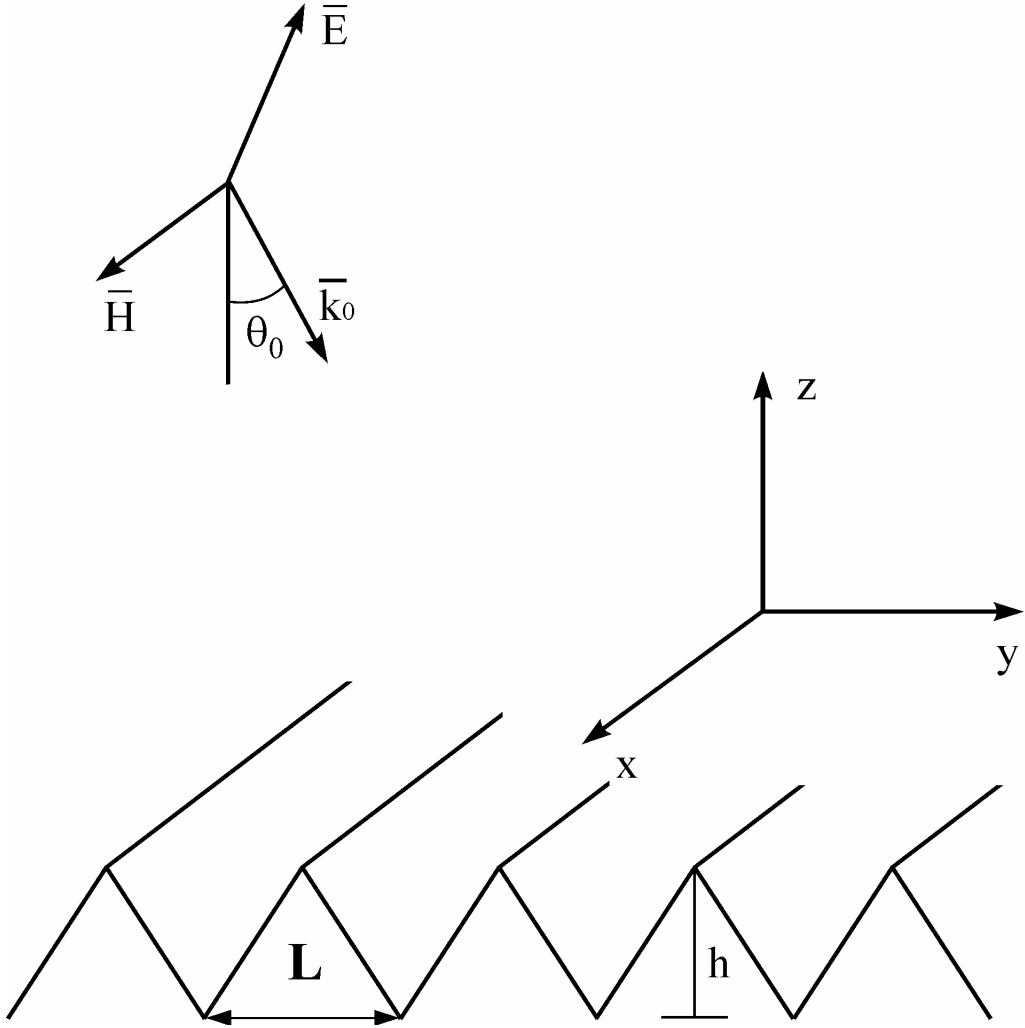

Figure 1. Regular lattice of a triangular profile $(L \ll \lambda)$. Here $L$ is the period of the lattice and $\lambda$ is the wavelength of the incident light, $h$ is the height of the lattice.

$$\overline{E}_{surf.sc.} = \sum_{n=-\infty}^{+\infty} \overline{g}_n e^{i\alpha_n y + i\gamma_n z} \tag{3}$$

$$\alpha_n = \frac{2\pi n + k_0 \sin\theta_0 L}{L} \tag{4}$$

$$\gamma_n = \sqrt{k_0^2 - \alpha_n^2} \tag{5}$$

Here $\overline{g}_n$ are the amplitudes, $n$ is the number of a spatial harmonic. For the period of the lattice $L \ll \lambda$ the spatial harmonic with $n=0$ is a direct reflected wave, while all the others are heterogeneous plane waves strongly localized near the surface. The maximum localization size has the harmonic with $n=1$. All the others are localized considerably stronger. The exact solution of the diffraction problem on the lattice reduces to determination of the coefficients $\overline{g}_n$. The main specific feature of the surface field is a steep or singular increase of the electric field near the wedges of the lattice or so-called rod effect. This type of behavior is independent on a particular surface profile. It is determined only by existence of sharp wedges. Besides it is independent on the dielectric properties of the lattice and exists in lattices with any dielectric constants that differ from the dielectric constant of vacuum. In the vicinity of the wedge (Figure 2a) the electric field can be estimated as

$$E_r = -g_{0,inc} C_0 \left(\frac{l_1}{r}\right)^\beta \sin(\lambda_1 \varphi)$$
$$E_\phi = -g_{0,inc} C_0 \left(\frac{l_1}{r}\right)^\beta \cos(\lambda_1 \varphi)$$
(6)

where $C_0$ is some numerical coefficient, ($l_1 = L$ or $h$) is a characteristic size of the lattice

$$\lambda_1 = \pi/(2\pi - \alpha) \tag{7}$$

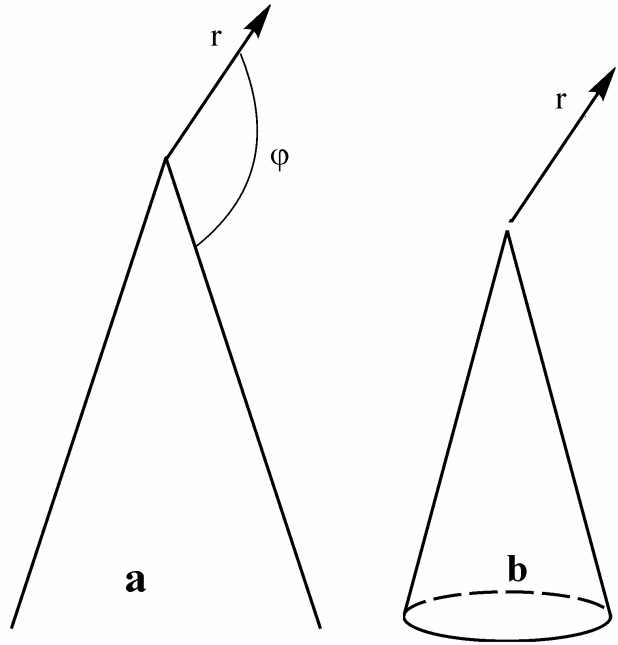

Figure 2. **a**-infinite wedge, **b**-the roughness of the cone type.

$\alpha$ is the wedge angle for an ideally conductive wedge.

$$\beta = 1 - \lambda_1 = \frac{\pi - \alpha}{2\pi - \alpha} \tag{8}$$

The specific feature of the field behavior (6) is appearance of the singularity $(l_1/r)^\beta$, which describes geometrical nature of the field enhancement. It determines the following behavior of the coefficients $\overline{g}_n$ in expression (3) [30]:

$$g_n \sim |n|^{\beta-1}. \tag{9}$$

Indeed, substitution of (9) into (3) gives

$$\sum_{\substack{n=-\infty \\ n \neq 0}}^{+\infty} |n|^{\beta-1} e^{2\pi |n| z/L} \sim 2 \int_0^\infty t^{\beta-1} e^{-2\pi z t/L} dt \sim 2\left(\frac{L}{2\pi z}\right)^\beta \tag{10}$$

For the wedge angles changing in the interval $0 < \alpha < \pi$ the $\beta$ value varies within the range $0 < \beta < 1/2$ and the coefficients $g_n$ slowly decrease as $n$ increases. Thus the singular behavior of the field arises because of specific summation of the surface waves at the top of the wedge. In the region of a three-dimensional roughness of the cone type (Figure 2b) the formula for estimation of the field has an approximate form

$$E_r \sim g_{0,inc} C_0 \left(\frac{l_1}{r}\right)^\beta, \tag{11}$$

where $\beta$ depends on the cone angle and varies within the interval $0 < \beta < 1$. Using formulae (6) and (11) one can note a very important property: a strong spatial variation of the field. For example

$$\frac{1}{E_r} \frac{\partial E_r}{\partial r} \sim \left(\frac{\beta}{r}\right) \tag{12}$$

can be significantly larger than the value $2\pi/\lambda$, which characterizes variation of the electric field in a free space. If one considers more realistic models of the rough surface than the regular metal lattice, it is obvious, that there is a strong enhancement of the perpendicular component of the electric field $\overline{E}_n$ at prominent places with a large curvature (Figure 3), while the tangential components $\overline{E}_\tau$ are comparable with the amplitude of the incident field. Besides, the electromagnetic field strongly varies in space with a characteristic length $l_E$ equal to characteristic roughness size. This type of behavior is not an exclusive property of the ideally conductive lattice and preserves near surfaces with a finite dielectric constant.

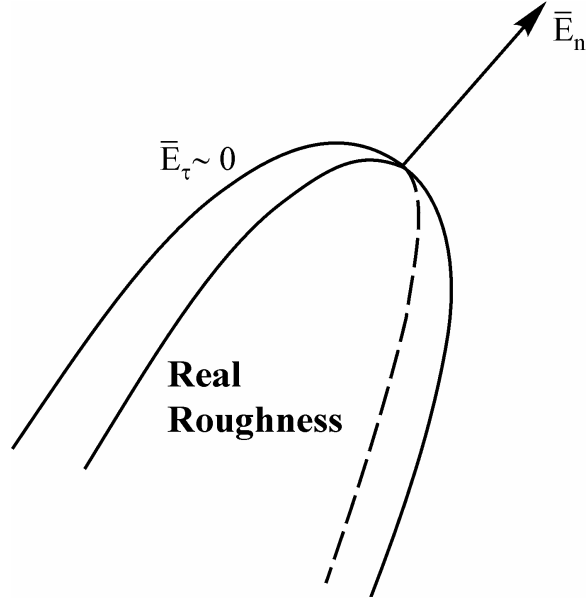

Figure 3. More realistic model of the roughness of the cone type. The normal component of the electric field $\overline{E}_n$ and the derivatives $\partial E_\alpha / \partial x_\alpha, \alpha = (x, y, z)$ are enhanced near the top of the roughness.

One should note that there is a number of works, which try to calculate the main features of the surface electromagnetic field near rough surface by numerical methods for various models of roughness. They are the discrete dipole approximation DDA method [31], the finite difference time-domain method (FDTD) [32, 33], the method of volume integral equation [34] and may be some others. The features of these methods one can find in the above references. Below we shall describe some results, obtained by these methods for some models of roughness and will compare them with the results obtained by our approach. It appears that the obtained results confirm strongly our point of view on the behavior of the field. In [32] the authors calculated the enhancement near some conical tip with round tip apex. The results qualitatively coincide with our ones. In fact there is a strong enhancement of the electric field in a narrow area of the round tip apex. In spite of the authors name this behavior as a manifestation of tip plasmons, this result do not differ from the behavior of the electric field due to the rod effect. Another series of works are the works of Kottmann et. al. [34-36] who calculated the behavior of the electric field near some models of nanowires using the numerical method of the volume integral equation [34]. In spite of the authors assign the enhancement to plasmon resonances, they measured the enhancement in the vicinity (1nm) of the wedges of hexagon, pentagon, square and triangle nanowires (Figure 4), where the rod effect is very strong.

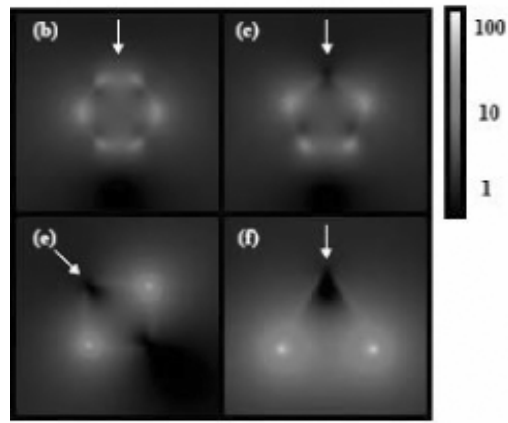

Figure 4. Diffraction of a plane electromagnetic wave on nanowires of hexagon, pentagon, square and triangle forms. The wave vector is perpendicular to the main axis of nanowires, while the $\bar{E}$ vector is in the plane of the figure [36]. One can see the strongest enhancement of the electric field in the vicinity of the wedges of nanowires that is manifestation of the rod effect.

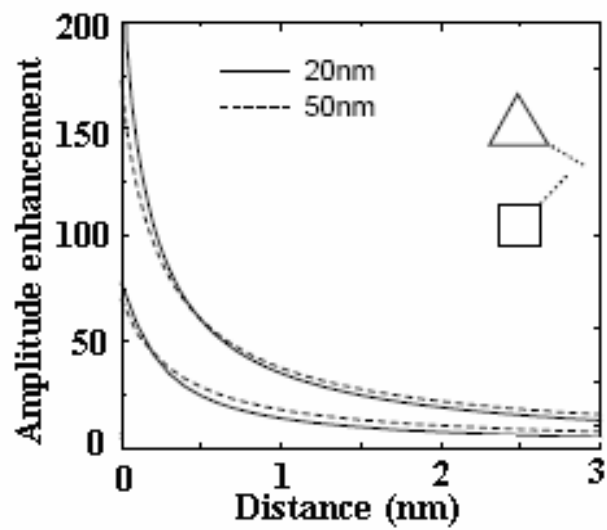

Figure 5. The distance dependences of the enhancement of the electric field from the wedges of nanowires of the square and triangle forms of 20 and 50 nm sizes [36]. The dependences qualitatively correspond to the behavior, described by formulae (6,11).

The measured distance dependence of the enhancement from the tops of the wedges (Figure 5) qualitatively confirms the dependences, described by formulae (6, 11). Because of the difference of the models in [35,36], and specificity of our and real models of rough surface, the calculated wavelength dependences of the enhancement (Figure 6 for example) does not reflect the main features of experimental wavelength dependences of SERS for arbitrary rough surfaces [7] (Figure 7 for example). The region of wavelengths of the resonances, calculated in [35, 36] is approximately (300-425) nm (Figure 6) with significantly smaller wavelengths than the region of the most enhancement in SERS on a real rough surface (450-700) nm (Figure 7). Taking into account, that the enhancement, caused by the nanowires can be measured at smaller distances from the wedges than 1 nm (as it was made in the calculations), the enhancement of the electric field can be significantly stronger than those obtained in [35,36].

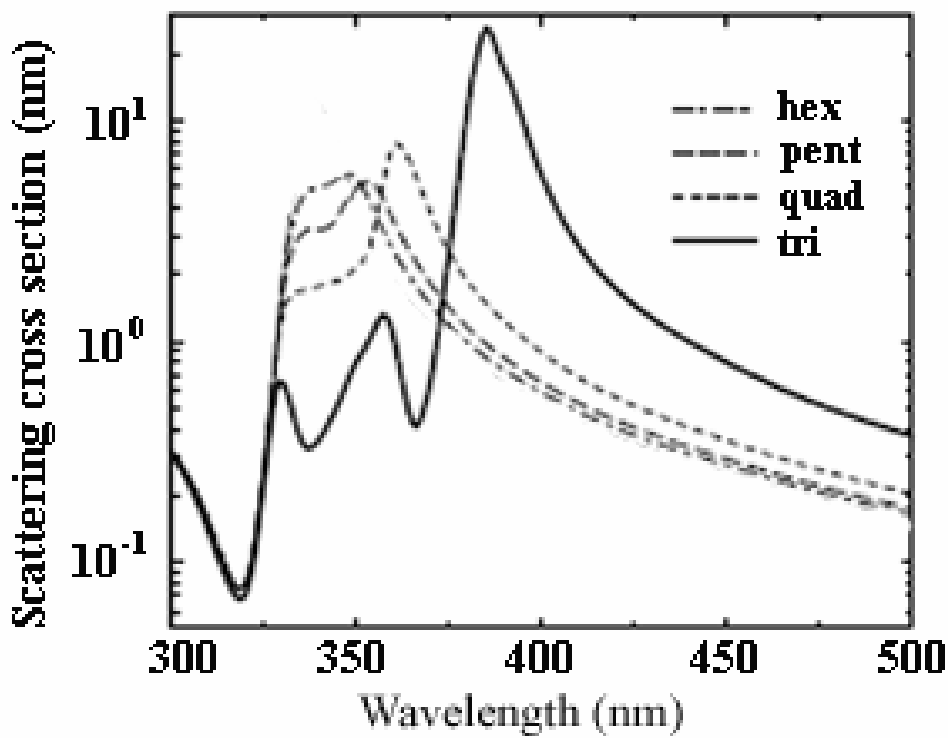

Figure 6. The wavelength dependences of the enhancement of the electric field near the wedges of nanowires of hexagon, pentagon, square and triangle forms [36]. One can see that the region of resonances corresponds to the wavelengths less than ~425nm.

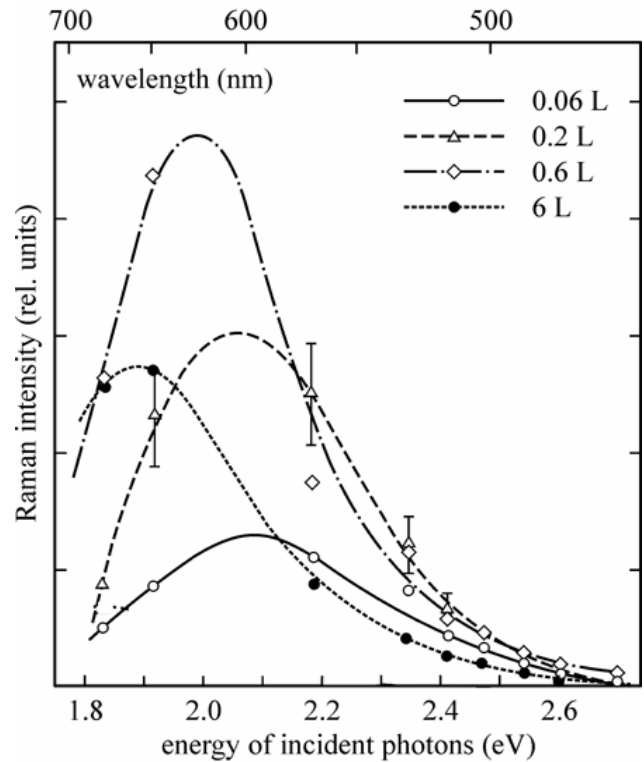

Figure 7. Typical wavelength dependences of the intensity of the SER bands for real rough surface of silver [7]. One can see that the region of the main enhancement corresponds to the wavelengths of the incident radiation $\lambda \geq 450$ nm.

Therefore all these results confirm the fact that the strongest enhancement, which arises in the vicinity of the wedges of the nanowires in the region of wavelengths, where the strong SERS is observed is caused not by the plasmon resonances but by the rod effect. One should note, that the detailed consideration of the derivatives of the electric field is absent in these papers such as in [32] and the real enhancement factor, which further is considered in our work is not considered in [35, 36].

As it was mentioned above there is calculation of the electric field between two close round nanowires, with parallel axis's [29]. It was established a strong enhancement of the electric field in the gap between them (Figure 8) and some resonances in the wavelength region (325-425) nm (Figure 9). However these resonances are situated in another wavelength range, than the range for maximum SERS (450-700) nm as in the previous case (Figure 7). Therefore this model can not explain the enhancement on arbitrary rough surface by these resonances too. This result is important from our point of view since it points out the existence of some additional active areas, or active sites (in A. Otto terminology [2]) where the strong enhancement of the electric field is possible.

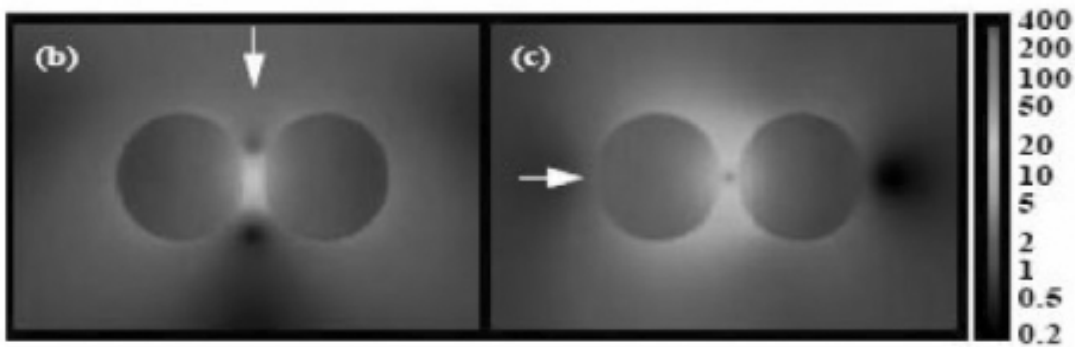

Figure 8. Diffraction of a plane electromagnetic wave on two coupled cylindrical nanowires of 50 nm. The incident direction is designated by arrows. The electric field is in the plane of the figure. One can see a region of the enhanced electric field in the gap between the nanowires [29].

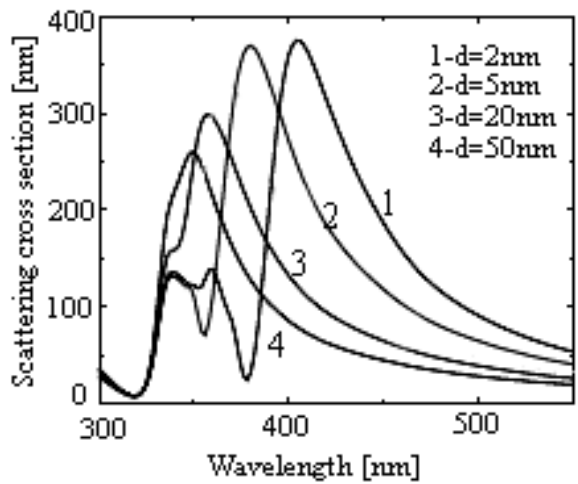

Figure 9. The wavelength dependence of the scattering cross-section for 50 nm coupled nanowires for various distances d between them [29]. One can see that the region of resonances for these models corresponds to the wavelengths $\lambda \leq$ (425-450) nm.

Analogous calculations by the finite-time domen method (FTDT) were presented in [37] for some chains of dimers. It was established that such areas of the enhancement of the electric field (active sites) or hot spots (in terminology of V. M. Shalaev) exist between two spheres of dimers too. However the strongest enhancement arises for the dimer consisting of tetrahedrons. The field enhancement is ~ $5.62 \times 10^2$ and ~$10^{11}$ for SERS for single dimer of two truncated tetrahedrons with 1 nm gap between their tops. These values can be increased in the lattice of dimers, using the long range effects, or space resonances, arising in their periodical chains. The field enhancement can achieve $1.77 \times 10^3$, while the enhancement for SERS ~$10^{13}$. The essential point of these investigations is that the maximum enhancement arises in the dimers with truncated tetrahedrons. Thus we again dealt with the rod effect in the vicinity of the truncated tops of the tetrahedrons and the main reason of this enhancement is the strong increase of the electric field near their tops, which are a good model of sharp roughness.

The ideal enhancement of the field at the top of the wedge, cone, tip or spike is infinity. The same result must be for the ideal sharp tops of tetrahedrons. However the existence of the truncation decreases strongly the SERS enhancement till $10^{13}$. One should note, that we achieved the SERS enhancement ~ $1.7 \times 10^{16}$ in our estimations for Single Molecule SERS [38] for the quadrupole enhancement mechanism. However one must take into account, that in all considered models we have some arbitrary parameters, which are defined from some reasonable physical point of view. Thus in spite of we use some another models of the roughness and their parameters than the authors of [37] the reason of the enhancement in all considered models is the same-the rod effect arising near sharp wedges or tips, or prominent places with a large curvature.

Thus the second type of the active sites, or hot spots arise in the gap between two parallel round nanowires or two spherical particles. The large enhancement can arise in the gap between two tops of tetrahedrons, cones, tips or spikes. The last cases are some limited cases of the existence of the enhancement in the gap between two nanowires [29], or between two spheres [37]. Moreover the enhancement in the last cases apparently can be the largest for sufficiently thin spikes with a small angle of the spike apex and a very small gap between them. In addition it must depend and be limited by their length, which must be equal or more than the wavelength of the incident field for the maximum enhancement. The last result follows from consideration of wave and electrostatic approximation. The analog of this enhancement geometry is two close spikes, situated on the same axis with a small gap between tops of them (Figure 10).

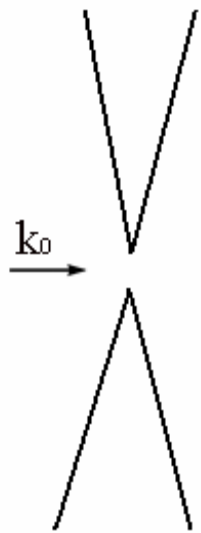

Figure 10. The limited case of the active sites of the second type. Apparently the most enhancement can be obtained in this geometry.

In this case they create electric arc between their tops under influence of the incident field. Apparently this geometry possesses by maximum enhancement.

Since for further consideration we use the main property of these models-the increase of the electric fields and their derivatives near the above features and qualitative consideration of the enhancement of the dipole and quadrupole light molecule interactions, the choice between these models is not of principle. Therefore further we shall use our models of the wedge or cone (spike), and formulae (6, 11) for consideration of the principal enhancement in SERS and SEHRS since it is sufficient for our goals.

## 3. Interaction of light with molecules near rough metal surface

In accordance with principles of theoretical physics the optical properties of molecules are determined by the light-molecule interaction Hamiltonian, which has the form

$$\hat{H}_{e-r} = -\sum_i \frac{ie\hbar}{mc} \bar{A}_i \nabla_i \tag{13}$$

Here the sign of e is positive. $\bar{A}_i$ is a vector potential of the electromagnetic field at the place of the $i$ electron. Other designations are conventional. For small objects, like molecules, the vector potential can be expanded in a Taylor series and final expressions for the light-molecule interaction Hamiltonians for the incident and scattered fields can be obtained in the form

$$\hat{H}^{inc}_{e-r} = |\bar{E}_{inc}| \frac{(\bar{e}^* \bar{f}_e^*)_{inc} e^{i\omega_{inc}t} + (\bar{e}\bar{f}_e)_{inc} e^{-i\omega_{inc}t}}{2} \tag{14}$$

$$\hat{H}^{scat}_{e-r} = |\bar{E}_{scat}| \frac{(\bar{e}^* \bar{f}_e^*)_{scat} e^{i\omega_{scat}t} + (\bar{e}\bar{f}_e)_{scat} e^{-i\omega_{scat}t}}{2} \tag{15}$$

where $\bar{E}_{inc}$ and $\bar{E}_{scat}$ are vectors of the incident and scattered electric fields, $\omega_{inc}$ and $\omega_{scat}$ are corresponding frequencies, $\bar{e}$ -are polarization vectors of the corresponding fields,

$$f_{e\alpha} = d_{e\alpha} + \frac{1}{2E_\alpha} \sum_\beta \frac{\partial E_\alpha}{\partial x_\beta} Q_{e\alpha\beta} \tag{16}$$

is an $\alpha$ component of the generalized vector of interaction of light with molecule,

$$d_{e\alpha} = \sum_i e x_{i\alpha} \qquad Q_{e\alpha\beta} = \sum_i e x_{i\alpha} x_{i\beta} \tag{17}$$

are the $\alpha$ component of the dipole moment vector and the $\alpha\beta$ component of the quadrupole moments tensor of interaction of light with electrons of the molecule. Here under $x_{i\alpha}$ and $x_{i\beta}$ we mean coordinates $x, y, z$ of $i$ electron. Usually the relative influence of the quadrupole and dipole light-molecule interactions is determined as the relation of the second and the first terms in the right side of (16). However this relation is the relation of quantum-mechanical operators, while all physical values are expressed via the matrix elements of these operators. Therefore one must consider, that the relative influence is determined by the relations

$$\frac{\langle m|Q_{e\alpha\beta}|n\rangle}{\langle m|d_{e\alpha}|n\rangle} \frac{1}{2E_\alpha} \frac{\partial E_\alpha}{\partial x_\beta} = B_{\alpha\beta} a \frac{1}{2E_\alpha} \frac{\partial E_\alpha}{\partial x_\beta} \tag{18}$$

where $a$ is a molecule size, $B_{\alpha\beta}$ are some numerical coefficients. The first factor in the left side of (18) is the relation of some mean matrix elements of the quadrupole and dipole transitions. It is necessary to point out that the $B_{\alpha\beta}$ values essentially differ for $\alpha \neq \beta$ and for $\alpha = \beta$. This results from the fact that $d_{e\alpha}$ and $Q_{e\alpha\beta}$ are the values with a changeable sign, while $Q_{e\alpha\alpha}$ are the values with a constant sign, that strongly increases the $B_{\alpha\alpha}$ values. The difficulty for this estimation is the fact, that there is no information about quadrupole transitions in molecules. Therefore because the inner shell configuration in molecules remains almost intact we usually take the value $\overline{\langle n|Q_{e\alpha\alpha}|n\rangle}$ instead of $\overline{\langle m|Q_{e\alpha\alpha}|n\rangle}$ and the value $\sqrt{e^2\hbar/2m\omega_{mn}} \times \sqrt{\overline{f}}_{mn}$ for $\overline{\langle m|d_\alpha|n\rangle}$, which is expressed in terms of some mean value of the oscillator strength $\overline{f}_{mn} = 0.1$ while $\omega_{mn}$ corresponds to the edge of absorption ($\lambda \approx 2500 A$). Since configuration of the electron shell is close to configuration of nuclei the value $\overline{\langle n|Q_{\alpha\alpha}|n\rangle}$ was calculated as a $Q_{n\alpha\alpha}$ component of the quadrupole moments of nuclei. Estimation for the pyridine, benzene or pyrazine molecules gives the value $B_{\alpha\alpha} \sim 2\times 10^2$ that strongly differs from the value $B_{\alpha\alpha} \sim 1$ ($B_{\alpha\alpha}a \sim a$). The last one is usually used in literature as the relation of the quadrupole and dipole operators. The full estimation of the relation (18) near rough metal surface is made for the models of roughness of the wedge or the cone or the spike types, which well approximate prominent features of the surface with a large curvature. Then using (12) one can obtain the following expression for

$$\frac{\overline{\langle m|Q_{e\alpha\alpha}|n\rangle}}{\overline{\langle m|d_{e\alpha}|n\rangle}} \frac{1}{2E_\alpha} \frac{\partial E_\alpha}{\partial x_\alpha} = B_{\alpha\alpha} a \frac{\beta}{r} \qquad (19)$$

It can be seen, that for $r < B_{\alpha\alpha} a \beta$ the quadrupole interaction can be more than the dipole one. The enhancement of the electric field and the dipole interaction can be estimated as

$$G_{H_d} \sim C_0 \left(\frac{l_1}{r}\right)^\beta \qquad (20)$$

while the enhancement of the quadrupole interaction compared to the dipole interaction in a free space can be estimated as

$$G_{H_Q} \sim C_0 \beta \left(\frac{B_{\alpha\alpha}}{2}\right)\left(\frac{l_1}{r}\right)^\beta \left(\frac{a}{r}\right) \qquad (21)$$

It can be seen, that for reasonable values $C_0 \sim 1$, $l_1 \sim 10$ nm, $r \sim 1$ nm, $\beta \sim 1$ and the molecules like pyridine, benzene or pyrazine with $B_{\alpha\alpha} \sim 2\times 10^2$ the enhancement of the dipole interaction is ~10, while the enhancement of the quadrupole interaction is ~ $10^2$.

## 4. The enhancement in SEHRS

The enhancement in SEHRS such as in SERS can be caused both by the enhancement of the $E_z$ component of the electric field, which is perpendicular to the surface and by the enhancement of the field derivatives and the quadrupole interaction. Since SEHRS is the process of the third order the enhancement due to the dipole interaction is

$$G_d \sim C_0^6 \left(\frac{l_1}{r}\right)^{6\beta} \qquad (22)$$

and

$$G_Q \sim C_0^6 \beta^6 \left(\frac{B_{\alpha\alpha}}{2}\right)^6 \left(\frac{l_1}{r}\right)^{6\beta} \left(\frac{a}{r}\right)^6 \qquad (23)$$

due to the quadrupole interaction. For the values of the parameters pointed out above, the enhancement due to the purely dipole interaction is of the order of $10^6$ while the enhancement due to the quadrupole interaction ~$10^{12}$. One should note that the enhancement of the dipole and quadrupole interactions $G_{H_d}$ and $G_{H_Q}$ and especially the enhancement of SEHRS $G_Q$ can be very large. For example for some limited situations with the values of the parameters $C_0 \sim 1$, $B_{\alpha\alpha} \sim 2\times 10^2$, $r \sim 0.1 nm$, $\beta \sim 1$, $l_1 \sim 100 nm$ corresponding to the placement of the molecule on the top of the cone (tip or spike) the enhancement $G_Q$ in SEHRS can achieve $10^{30}$. As it was mentioned above the real situation is that most enhancement arises in the vicinity of some points associated with prominent places with very large curvature. The mean enhancement is formed from the whole layer of adsorbed molecules and is significantly smaller than the maximum enhancement near these places due to averaging.

## 5. Main and minor moments

In accordance with the above estimations one can introduce a conception of main and minor moments. The main dipole and quadrupole moments are those, which are responsible for the strong enhancement. This classification essentially depends on orientation of the molecules near rough surface. Usually the experiments are performed on molecules in solutions and the substrate is under some negative applied potential. As it is well known, the strongest enhancement arises from molecules situated directly at the surface in the first layer and the enhancement strongly decreases with the increase of the distance between the molecule and the surface. This effect arises from the strong decrease of the electric field and its derivatives with the distance [8-10]. Further we shall consider the pyrazine and phenazine molecules and designate the coordinate system, which is associated with the molecules as $(x, y, z)$ and the one with the surface as $(x', y', z')$. We consider that the plane of the molecules coincide with the $XZ$ plane and the $z$ axis passes through nitrogen atoms. The $z'$ axis is perpendicular to the surface. The molecules can bind with the surface via the nitrogen lone pair having the end on orientation. Other molecules are in the solution and can have an arbitrary orientation. The number of molecules, which are bound with the surface apparently is large, because of the binding, which prevents their movement and arbitrary orientation in the solution. The negative potential prevents the binding. Therefore the number of the bound molecules apparently depends on the potential. The number of arbitrary oriented molecules strongly depends on their concentration in the solution. Apparently their number situated directly at the surface decreases with the increasing of the negative potential, because the binding becomes less energetically favorable and more molecules at the surface become free. Because of existence of molecules which are oriented arbitrary with respect to the surface and hence to the enhanced $E_{z'}$ component of the electric field all the $d_\alpha$ and $Q_{\alpha\alpha}$ quadrupole moments are essential for the scattering and are the main ones. As it is well demonstrated in [10], the other $Q_{xy}, Q_{xz}, Q_{yz}$ quadrupole moments are not essential, because of their changeable sign and we name them as minor moments. In order to receive an idea about the role of the dipole and quadrupole interactions in SEHRS it is reasonable to consider the SEHR spectra of some symmetrical molecules, that can give precise information about allowed and forbidden bands.

Since our further consideration concerns symmetrical molecules and SEHRS selection rules in these molecules, it is necessary to determine the minor and the main moments for this case. It is convenient to transfer to the values, which transform after irreducible representations of the symmetry group. Analysis of the tables of irreducible representations of all point groups [10, 39] demonstrates that all the $d$ and $Q_{xy}, Q_{xz}, Q_{yz}$ moments transform after irreducible representations of the major part of the point symmetry groups, while the $Q_{xx}, Q_{yy}, Q_{zz}$ moments can be expressed via linear combinations $Q_1, Q_2, Q_3$ transforming after irreducible representations.

$$
\begin{aligned}
Q_{xx} &= a_{11}Q_1 + a_{12}Q_2 + a_{13}Q_3 \\
Q_{yy} &= a_{21}Q_1 + a_{22}Q_2 + a_{23}Q_3 \\
Q_{zz} &= a_{31}Q_1 + a_{32}Q_2 + a_{33}Q_3
\end{aligned}
\qquad (24)
$$

where the coefficients $a_{ij}$ depend on the symmetry group. The corresponding expressions for $Q_1, Q_2, Q_3$ are

$$
\begin{aligned}
Q_1 &= b_{11}Q_{xx} + b_{12}Q_{yy} + b_{13}Q_{zz} \\
Q_2 &= b_{21}Q_{xx} + b_{22}Q_{yy} + b_{23}Q_{zz} \\
Q_3 &= b_{31}Q_{xx} + b_{32}Q_{yy} + b_{33}Q_{zz}
\end{aligned}
\qquad (25)
$$

Here the coefficients $b_{ij}$ depend on the symmetry group too. The specific form of $Q_1, Q_2, Q_3$ for some point groups one can find in [10] and in Appendix. There are combinations with a constant sign, which we can name as the main moments and of a changeable sign, which can be named as the minor ones. In accordance with our previous consideration the main moments are responsible for the strong enhancement, while the minor moments are nonessential for the scattering. For phenazine and pyrazine molecules, which are considered in this paper $Q_1, Q_2$ and $Q_3$ coincide with $Q_{xx}, Q_{yy}$ and $Q_{zz}$

## 6. SEHRS in symmetrical molecules

The above estimations give only understanding of the principle possibility of the SEHRS enhancement mechanism. A good prove of its validity can be obtained from consideration of selection rules and regularities of the SEHR spectra of symmetrical molecules. The detailed mathematical consideration of SEHRS is very cumbersome. However one can receive main results from some physical point of view on the base of the methods used in the SERS theory [10]. SEHRS is a three photon process, which occurs via the dipole and quadrupole moments. In case we consider the wavefunctions of symmetrical molecules transforming after irreducible representations of the molecule symmetry group, the full cross-section of the $s$ vibrational band can be expressed as a sum of various scattering contributions, which occur via various combinations of the dipole and quadrupole moments such as in SERS (formula 59 in [10]).

$$\sigma_{SEHRS_s} \sim \sum_p \binom{(V_{(s,p)}+1)/2}{V_{(s,p)}/2} \left| \sum_{f_1, f_2, f_3} T_{(s,p), f_1-f_2-f_3} \right|^2 \qquad (26)$$

Here $f_1, f_2, f_3$ are various dipole and quadrupole moments, transforming after irreducible representations of the symmetry group of the molecule. The values of contributions $T_{(s,p), f_1-f_2-f_3}$ are expressed via the values

$$R_{n,l,(s,p)} \langle n|f_1|m\rangle\langle m|f_2|k\rangle\langle k|f_3|l\rangle \qquad (27)$$

and similar expressions, with various permutations of the moments. Here $(s, p)$ designates the quantum numbers of degenerated vibrations of the molecule. $s$ numerates the group of degenerated vibrations, while $p$ numerates the vibrations inside the group. $n$ designates the ground, while $l, k, m$ the excited states of the molecule. $R_{n,l,(s,p)}$ are the coefficients of excitation of the $l$ electronic states due to vibration $(s, p)$. Their evident form can be found in [10, 17, 24]. In accordance with consideration, similar to [10, 17, 24], the following selection rules can be obtained

$$\Gamma_{(s,p)} \in \Gamma_{f_1} \Gamma_{f_2} \Gamma_{f_3} \tag{28}$$

where the symbol $\Gamma$ designates the irreducible representation of the corresponding moment $f$ and of the vibration $(s, p)$. The physical sense of the expressions of the type (27) is that the contributions are expressed via the sequence of quantum transitions arising via various dipole and quadrupole moments and their permutations. In accordance with our previous consideration the most enhancement for the strongly rough surface is caused by the quadrupole interaction with $Q_{main}$ moments and by the dipole interaction with the main dipole moments $d_{main}$. Then the contributions $T_{(s,p), f_1 - f_2 - f_3}$ which we shall designate further simply as $(f_1 - f_2 - f_3)$ can be classified qualitatively after the enhancement degree in the following manner:

1. $(Q_{main} - Q_{main} - Q_{main})$ - the most enhanced scattering type.

2. $(Q_{main} - Q_{main} - d_{main})$ - scattering type, which can be strongly enhanced too, but in a lesser degree than the previous one.

3. $(Q_{main} - d_{main} - d_{main})$ - scattering type, which can be strongly enhanced too, but lesser, than the two previous ones and

4. $(d_{main} - d_{main} - d_{main})$ - scattering type, which can be strongly enhanced too, but lesser than the three previous ones.

Here and further we mean under $(f_1 - f_2 - f_3)$ all contributions with permutations of the $f$ moments. Considering molecules with the groups, where all $d_{main}$ moments do not transfer after the unit representation, one can note, that the first and the third enhancement types with the same $d_{main}$ moments contribute to the bands, caused by vibrations transforming after the unit irreducible representation or by the totally symmetric vibrations. The second and types contribute to the bands, caused by vibrations transforming such as the $d_{main}$ moment. Thus the most enhanced bands in the molecules with the pointed groups are caused by the above types of vibrations. The first ones are forbidden in usual HRS. Thus these bands must be an essential feature of the SEHR spectra of symmetrical molecules with the above symmetry. The contributions of ( $Q_{main} - d_{main} - d_{main}$ ) with various $d_{main}$ and ( $d_{main} - d_{main} - d_{main}$ ) scattering types can manifest in the SEHR spectra. The contributions which contain $Q_{\min or}$ moments apparently will be small and may manifest in the SEHR spectra only under very strong incident field.

## 7. Analysis of the SEHR spectrum of phenazine

The amount of the works, which analyze the SEHR spectra of symmetrical molecules is very small, such as the number of works on other aspects of SEHRS. Here we shall analyze some previous results and peculiarities of the SEHR spectra of phenazine [40] and pyrazine [21, 22], published and interpreted by another authors. The main feature of these works is consideration of the SEHR spectra using the dipole approximation of the light-molecule interaction Hamiltonian only. Let us consider the results of measurement of the SEHR spectrum, which refer to phenazine [40]. The SEHR spectra of this molecule are presented on the Figure 11. The main difficulty in this matter is assignment of vibrations to various irreducible representations of the $D_{2h}$ symmetry, which describe symmetry of this molecule. A good assignment one can find in [41]. However several modes which may belong to various irreducible representations are very close. Therefore sometime we are not able to assign them unequivocally. The main feature of the phenazine SEHR spectra both for 0 V and - 0.2 V applied potentials is appearance of sufficiently strong bands with $A_g$ symmetry which are forbidden in usual HRS. They are the bands with 416, 1015, 1166, 1406 and 1560 $cm^{-1}$ (Table 1). Here we write out the wavenumbers, which refer to the 0 V applied potential. In accordance with our ideas the enhancement of these bands arise due to the strong dipole and quadrupole light-molecule interactions and is caused by the ($Q_{main} - Q_{main} - Q_{main}$) types of the scattering, and by ($Q_{main} - d_{main} - d_{main}$) types of the scattering with the same $d_{main}$ moments. The bands at 1277 $cm^{-1}$ and 1475 $cm^{-1}$ may refer both to the $A_g$ and $B_{1u}$ symmetry types, because of uncertainty of rigorous assignment of these bands associated with close values of the wavenumbers used for the assignment [41]. However, in both cases they can be explained by the ($Q_{main} - Q_{main} - Q_{main}$) and ($Q_{main} - d_{main} - d_{main}$) scattering types for the $A_g$ type of symmetry and by $(Q_{main} - Q_{main} - d_z)$ and $(d_{main} - d_{main} - d_z)$ contributions with the same $d_{main}$ moments of arbitrary oriented molecules for the $B_{1u}$ type of symmetry. Both types of the lines should be strongly enhanced and manifest in the SEHR spectrum of this molecule. The band at 754 $cm^{-1}$ of $B_{2u}$ symmetry type can be explained by the $(Q_{main} - Q_{main} - d_y)$ and $(d_{main} - d_{main} - d_y)$ scattering types of arbitrary oriented molecules with the same $d_{main}$ moments. Thus existence of all the bands of phenazine observed in [40] can be explained by the dipole-quadrupole theory.

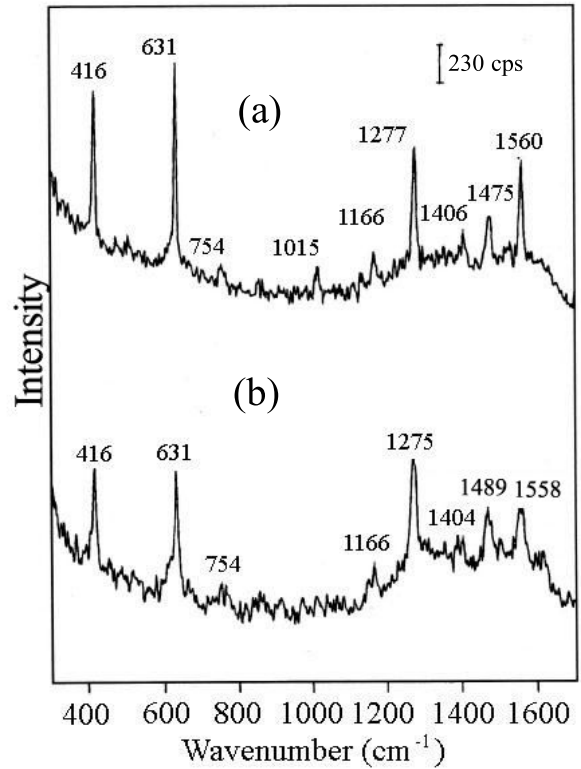

Figure 11. The SEHR spectra of phenazine adsorbed on silver electrode in a solution of saturated phenazine (~ $10^{-4}$ M) + 0.01 M $HClO_4$ + 0.1 M KCl at the potentials (a) 0 V and (b) -0.2 V. Appearance of the bands at 416, 1015, 1166, 1406 and 1560 $cm^{-1}$ with $A_g$ irreducible representation strongly confirms the dipole-quadrupole SEHRS theory.

Table 1. Symmetry and the wavenumbers of observable bands of the SEHR spectrum of phenazine. The values of wavenumbers in parenthesis correspond to those in [41].

| SEHRS | | Assignment | | | |
|---|---|---|---|---|---|
| 0 V $(cm^{-1})$ | -0.2 V $(cm^{-1})$ | $A_g$ | $B_{1u}$ | $B_{2u}$ | $B_{3u}$ |
| 416s | 416s | $A_g$ | | | |
| 631vs | 631vs | | $B_{1u}$ (657) | | |
| 754w | 754vw | | | $B_{2u}$ (749) | |
| 1015w | | $A_g$ | | | |
| 1166w | 1166w | $A_g$ | | | |
| 1277s | 1275s | $A_g$ (1280) | $B_{1u}$ (1275) | | |
| 1406w | 1404w | $A_g$ | | | |
| 1475m | 1469m | $A_g$ (1479) | $B_{1u}$ (1471) | | |
| 1560s | 1558m | $A_g$ | | | |

## 8. Analysis of the SEHR spectrum of pyrazine

Analysis of experimental SEHR spectrum of pyrazine obtained in [21, 22] (Figure 12) also confirms the dipole-quadrupole theory because it allows also to explain appearance of the strong forbidden bands with $A_g$ symmetry. The SEHR spectra, obtained in [21, 22] slightly differ one from another because of some difference in experimental conditions. However the bands symmetry is well known that is sufficient for analysis of the SEHR spectra. Below we shall analyze the SEHR spectrum obtained in [21]. In accordance with the selection rules (28) the lines caused by the vibrations with the following symmetry are observed in the SEHR spectrum (Table 2).

1. $A_g$ - (636, 1021, 1228, 1593 and 1624 $cm^{-1}$) caused mainly by ($Q_{main} - Q_{main} - Q_{main}$) and ($Q_{main} - d_{main} - d_{main}$), scattering contributions with the same $d_{main}$ moments of arbitrary oriented molecules. These lines are forbidden in usual HRS and their appearance strongly proves our point of view.

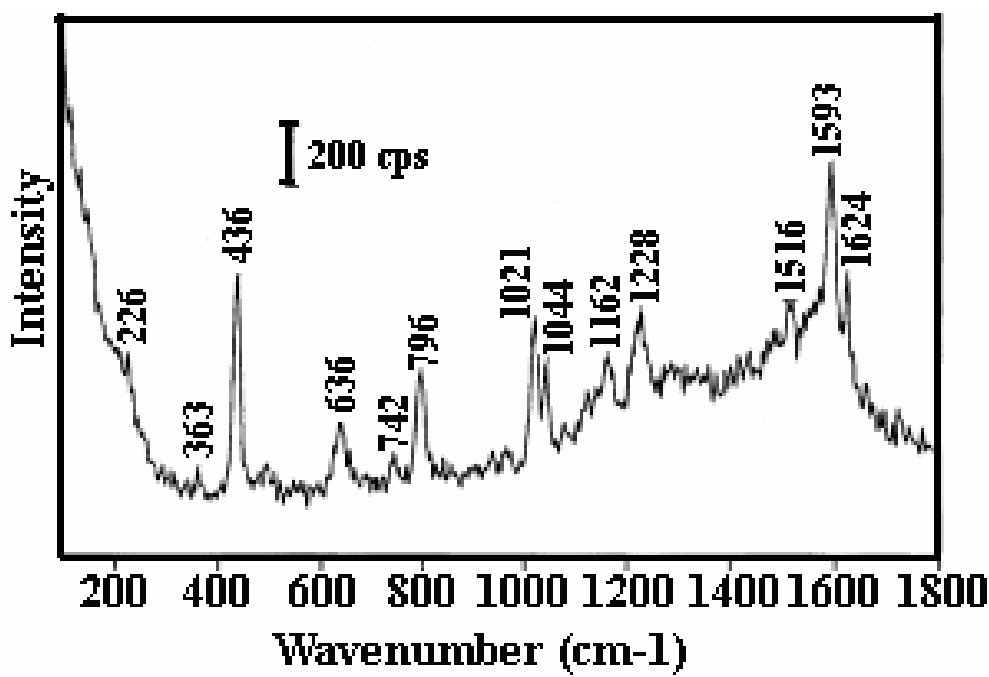

Figure 12. The SEHR spectrum of pyrazine [21]. One can see appearance of the strong bands of $A_g$ symmetry at 636, 1021, 1228, 1593 and 1624 $cm^{-1}$, arising due to the strong quadrupole light-molecule interaction.

Table 2  Symmetry and the wavenumbers of observable bands of the SEHR spectrum of pyrazine.

| Symmetry type | Wavenumbers for SEHRS ($cm^{-1}$) | Relative intensity |
|---|---|---|
| $A_g$ | 1624 | m |
| $A_g$ | 1593 | vs |
| $A_g$ | 1228 | s |
| $A_g$ | 1021 | s |
| $A_g$ | 636 | s |
| $B_{1u}$ | 1044 | m |
| $B_{2u}$ | 796 | s |
| $B_{2u}$ | 436 | vs |
| $B_{3u}$ | 1162 | m |
| $B_{2g}$ | 1516 | w |
| $B_{3g}$ | 742 | w |

2. $B_{1u}$ - 1044 $cm^{-1}$ caused mainly by the $(Q_{main} - Q_{main} - d_z)$ and $(d_{main} - d_{main} - d_z)$ scattering contributions with the same $d_{main}$ moments of vertically and arbitrary oriented pyrazine. Apparently the contributions $(Q_{main} - Q_{main} - d_z)$ and $(d_y - d_y - d_z)$ of horizontally adsorbed pyrazine are small since they contain the $d_z$ moment, which is associated with the non enhanced $E_{y'}$ tangential component of the electric field for this orientation. Thus the enhancement of the band with $B_{1u}$ symmetry is caused mainly by the above contributions.

3. $B_{2u} - 436$ and 796 $cm^{-1}$ caused mainly by the $(Q_{main} - Q_{main} - d_y)$ and $(d_{main} - d_{main} - d_y)$ scattering contributions with the same $d_{main}$ moments of horizontally and arbitrary oriented molecules. The vertically adsorbed pyrazine apparently does not determine the intensity of the bands with $B_{2u}$ symmetry, since the corresponding contributions include the non enhanced tangential $E_{y'}$ component of the electric field. Thus the enhancement of the bands with $B_{2u}$ symmetry is caused mainly by the above contributions of horizontally and arbitrary oriented molecules.

4. $B_{3u}$ - 1162 $cm^{-1}$ caused mainly by the ($Q_{main} - Q_{main} - d_x$) and ($d_{main} - d_{main} - d_x$) scattering contributions with the same $d_{main}$ moments of arbitrary oriented pyrazine.

5. The lines of the $B_{2g}$ and $B_{3g}$ symmetry (1516 and 742 $cm^{-1}$ respectively) may be caused mainly by ($Q_{main} - d_z - d_x$) and ($Q_{main} - d_z - d_y$) of arbitrary oriented molecules. Apparently the vertically and horizontally oriented molecules do not determine the intensities of the bands with $B_{2g}$ and $B_{3g}$ symmetry because of the presence of the non enhanced tangential components of the electric field in corresponding contributions.

The absence of the lines of $A_u$ and $B_{1g}$ symmetry caused by ($d_z - d_x - d_y$). ($Q_{main} - d_x - d_y$) and similar scattering contributions may be caused by small enhancement of these lines in the spectrum due to the large number of the components of the electric field, which are not enhanced significantly, that determine not significant value of the enhancement of the above contributions. Thus appearance of all the lines in the SEHR spectrum of pyrazine can be successfully explained by our theory, while appearance of the strong lines with $A_g$ symmetry strongly confirms existence of the strong quadrupole light-molecule interaction.

## 9. Conclusion

Thus the analysis of electromagnetic fields for several models of rough surfaces and separate models of roughness points out the existence of a very large enhancement in vicinity of the tops of such models of roughness as wedges, cones, tips or spikes. For more real models of roughness the most enhancement arises in the narrow areas of the tops of the prominent places with a large curvature. Analogous areas arise between two nanowires, or two spheres, when the gap between them is very small. The limited case for these situations is the particles with sharp points like spikes. Apparently it is the most preferable configuration in order to obtain the maximum enhancement of the electric field. One should note that there will be not only the strong enhancement of the electric field, but its derivatives too. The last fact and the quantum mechanical features of the quadrupole light-molecule interaction cause the enhancement both of the dipole and quadrupole light-molecule interactions. However the most fruitful approach to the analysis of this mechanism is investigation of the SER and SEHR spectra of symmetrical molecules. Analysis of the appearance of forbidden and allowed lines in these molecules gives very precise information about the validity of this mechanism. Specific consideration of the SEHR spectra of phenazine and pyrazine allows to explain appearance of the new SEHRS bands caused by vibrations transforming after the unit irreducible representation of the $D_{2h}$ symmetry group, which are forbidden in the pure dipole and are allowed in the dipole-quadrupole theory that strongly support the last theory. More detailed mathematical exposition of this theory will be published elsewhere.

## 10. Appendix

Table 3. The table of irreducible representations of the $D_{2h}$ symmetry group

| $D_{2h}$ | $C_1$ | $\sigma(xy)$ | $\sigma(xz)$ | $\sigma(yz)$ | $I$ | $C_2(z)$ | $C_2(y)$ | $C_2(x)$ | The dipole and quadrupole moments |
|---|---|---|---|---|---|---|---|---|---|
| $A_g$ | 1 | 1 | 1 | 1 | 1 | 1 | 1 | 1 | $Q_{xx}, Q_{yy}, Q_{zz},$ |
| $A_u$ | 1 | -1 | -1 | -1 | -1 | 1 | 1 | 1 | |
| $B_{1g}$ | 1 | 1 | -1 | -1 | 1 | 1 | -1 | -1 | $Q_{xy}$ |
| $B_{1u}$ | 1 | -1 | 1 | 1 | -1 | 1 | -1 | -1 | $d_z$ |
| $B_{2g}$ | 1 | -1 | 1 | -1 | 1 | -1 | 1 | -1 | $Q_{xz}$ |
| $B_{2u}$ | 1 | 1 | -1 | 1 | -1 | -1 | 1 | -1 | $d_y$ |
| $B_{3g}$ | 1 | -1 | -1 | 1 | 1 | -1 | -1 | 1 | $Q_{yz}$ |
| $B_{3u}$ | 1 | 1 | 1 | -1 | -1 | -1 | -1 | 1 | $d_x$ |

The main quadrupole moments in this group are $Q_1 = Q_{xx}, Q_2 = Q_{yy}$ and $Q_3 = Q_{zz}$